# Dimensions of Copeland-Erdös Sequences

Xiaoyang Gu,* Jack H. Lutz,* and Philippe Moser†
Department of Computer Science
Iowa State University
Ames, IA 50011, USA
{xiaoyang,lutz,moser}@cs.iastate.edu

**Abstract**

The base-$k$ *Copeland-Erdös sequence* given by an infinite set $A$ of positive integers is the infinite sequence $\mathrm{CE}_k(A)$ formed by concatenating the base-$k$ representations of the elements of $A$ in numerical order. This paper concerns the following four quantities.

- The *finite-state dimension* $\dim_{\mathrm{FS}}(\mathrm{CE}_k(A))$, a finite-state version of classical Hausdorff dimension introduced in 2001.

- The *finite-state strong dimension* $\mathrm{Dim}_{\mathrm{FS}}(\mathrm{CE}_k(A))$, a finite-state version of classical packing dimension introduced in 2004. This is a dual of $\dim_{\mathrm{FS}}(\mathrm{CE}_k(A))$ satisfying $\mathrm{Dim}_{\mathrm{FS}}(\mathrm{CE}_k(A)) \geq \dim_{\mathrm{FS}}(\mathrm{CE}_k(A))$.

- The *zeta-dimension* $\mathrm{Dim}_\zeta(A)$, a kind of discrete fractal dimension discovered many times over the past few decades.

- The *lower zeta-dimension* $\dim_\zeta(A)$, a dual of $\mathrm{Dim}_\zeta(A)$ satisfying $\dim_\zeta(A) \leq \mathrm{Dim}_\zeta(A)$.

We prove the following.

1. $\dim_{\mathrm{FS}}(\mathrm{CE}_k(A)) \geq \dim_\zeta(A)$. This extends the 1946 proof by Copeland and Erdös that the sequence $\mathrm{CE}_k(\mathrm{PRIMES})$ is Borel normal.

2. $\mathrm{Dim}_{\mathrm{FS}}(\mathrm{CE}_k(A)) \geq \mathrm{Dim}_\zeta(A)$.

3. These bounds are tight in the strong sense that these four quantities can have (simultaneously) any four values in $[0, 1]$ satisfying the four above-mentioned inequalities.

## 1 Introduction

In the early years of the twenty-first century, two quantities have emerged as robust, well-behaved, asymptotic measures of the finite-state information content of a given sequence $S$ over a finite alphabet $\Sigma$. These two quantities, the *finite-state dimension* $\dim_{\mathrm{FS}}(S)$ and the *finite-state strong dimension* $\mathrm{Dim}_{\mathrm{FS}}(S)$ (defined precisely in section 3), are duals of one another satisfying $0 \leq \dim_{\mathrm{FS}}(S) \leq \mathrm{Dim}_{\mathrm{FS}}(S) \leq 1$ for all $S$. They are mathematically well-behaved, because they are natural effectivizations of the two most important notions of fractal dimension. Specifically, finite-state dimension is a finite-state version of classical Hausdorff dimension introduced by Dai, Lathrop, Lutz, and Mayordomo [10], while finite-state strong dimension is a finite-state version of classical packing dimension introduced by Athreya, Hitchcock, Lutz, and Mayordomo [3]. Both finite-state dimensions, $\dim_{\text{FS}}(S)$ and $\text{Dim}_{\text{FS}}(S)$, are robust in that each has been exactly characterized in terms of finite-state gamblers [10, 3], information-lossless finite-state compressors [10, 3], block-entropy rates [5], and finite-state predictors in the log-loss model [14, 3]. In each case, the characterizations of $\dim_{\text{FS}}(S)$ and $\text{Dim}_{\text{FS}}(S)$ are exactly dual, differing only in that a limit inferior appears in one characterization where a limit superior appears in the other. Hence, whether we think of finite-state information in terms of gambling, data compression, block entropy, or prediction, $\dim_{\text{FS}}(S)$ and $\text{Dim}_{\text{FS}}(S)$ are the lower and upper asymptotic information contents of $S$, as perceived by finite-state automata.

---

*This research was supported in part by National Science Foundation Grant 0344187.
†This research was supported in part by Swiss National Science Foundation Grant PBGE2–104820.

For any of the dimensions mentioned above, whether classical or finite-state, calculating the dimension of a particular object usually involves separate upper and lower bound arguments, with the lower bound typically more difficult. For example, establishing that $\dim_{\text{FS}}(S) = \alpha$ for some particular sequence $S$ and $\alpha \in (0,1)$ usually involves separate proofs that $\alpha$ is an upper bound and a lower bound for $\dim_{\text{FS}}(S)$. The upper bound argument, usually carried out by exhibiting a particular finite-state gambler (or predictor, or compressor) that performs well on $S$, is typically straightforward. On the other hand, the lower bound argument, proving that *no* finite-state gambler (or predictor, or compressor) can perform better on $S$, is typically more involved.

This paper exhibits and analyzes a flexible method for constructing sequences satisfying given lower bounds on $\dim_{\text{FS}}(S)$ and/or $\text{Dim}_{\text{FS}}(S)$. The method is directly motivated by work in the first half of the twentieth century on Borel normal numbers. We now review the relevant aspects of this work.

In 1909, Borel [4] defined a sequence $S$ over a finite alphabet $\Sigma$ to be *normal* if, for every string $w \in \Sigma^+$,
$$\lim_{n \to \infty} \frac{1}{n} |\{ i < n \mid S[i..i + |w| - 1] = w \}| = |\Sigma|^{-|w|},$$
where $S[i..j]$ is the string consisting of the $i$th through $j$th symbols in $S$. That is, $S$ is normal (now also called *Borel normal*) if all the strings of each length appear equally often, asymptotically, in $S$. (Note: Borel was interested in numbers, not sequences, and defined a real number to be *normal in base $k$* if its base-$k$ expansion is normal in the above sense. Subsequent authors mentioned here also stated their results in terms of real numbers, but we systematically restate their work in terms of sequences.)

The first explicit example of a normal sequence was produced in 1933 by Champernowne [7], who proved that the sequence
$$S = 123456789101112\cdots, \tag{1.1}$$
formed by concatenating the decimal expansions of the positive integers in order, is normal over the alphabet of decimal digits. Of course there is nothing special about decimal here, i.e., Champernowne's argument proves that, for any $k \geq 2$, the sequence (now called the base-$k$ *Champernowne sequence*) formed by concatenating the base-$k$ expansions of the positive integers in order is normal over the alphabet $\Sigma_k = \{0, 1, \ldots, k-1\}$.

Champernowne [7] conjectured that the sequence
$$S = 235711131719232931\cdots, \tag{1.2}$$
formed by concatenating the decimal expansions of the prime numbers in order, is also normal. Copeland and Erdös [8] proved this conjecture in 1946, and it is the method of their proof that is of interest here. Given an infinite set $A$ of positive integers and an integer $k \geq 2$, define the base-$k$

*Copeland-Erdös sequence* of $A$ to be the sequence $\text{CE}_k(A)$ over the alphabet $\Sigma_k = \{0, 1, \ldots, k-1\}$ formed by concatenating the base-$k$ expansions of the elements of $A$ in order. The sequences (1.1) and (1.2) are thus $\text{CE}_{10}(\mathbb{Z}^+)$ and $\text{CE}_{10}(\text{PRIMES})$, respectively, where $\mathbb{Z}^+$ is the set of all positive integers and PRIMES is the set of prime numbers. Say that a set $A \subseteq \mathbb{Z}^+$ satisfies the *Copeland-Erdös hypothesis* if, for every real number $\alpha < 1$, for all sufficiently large $n \in \mathbb{Z}^+$,

$$|A \cap \{1, 2, \ldots, n\}| > n^\alpha.$$

Copeland and Erdös [8] proved that *every* set $A \subseteq \mathbb{Z}^+$ satisfying the Copeland-Erdös hypothesis has the property that, for *every* $k \geq 2$, the sequence $\text{CE}_k(A)$ is normal over the alphabet $\Sigma_k$. The normality of the sequence (1.2) – and of all the sequences $\text{CE}_k(\text{PRIMES})$ – follows immediately by the Prime Number Theorem [1, 13], which says that

$$\lim_{n \to \infty} \frac{|\text{PRIMES} \cap \{1, 2, \ldots, n\}| \ln n}{n} = 1,$$

whence PRIMES certainly satisfies the Copeland-Erdös hypothesis.

The significance of the Copeland-Erdös result for finite-state dimension lies in the fact that the Borel normal sequences are known to be precisely those sequences that have finite-state dimension 1 [16, 5]. The Copeland-Erdös result thus says that the sequences $\text{CE}_k(A)$ have finite-state dimension 1, provided only that $A$ is "sufficiently dense" (i.e., satisfies the Copeland-Erdös hypothesis).

In this paper, we generalize the Copeland-Erdös result by showing that a parametrized version of the Copeland-Erdös hypothesis for $A$ gives lower bounds on the finite-state dimension of $\text{CE}_k(A)$ that vary continuously with – in fact, coincide with – the parameter. The parametrization that achieves this is a quantitative measure of the asymptotic density of $A$ that has been discovered several times by researchers in various areas over the past few decades. Specifically, define the *zeta-dimension* of a set $A \subseteq \mathbb{Z}^+$ to be

$$\text{Dim}_\zeta(A) = \inf\{s \mid \zeta_A(s) < \infty\},$$

where the *$A$-zeta function* $\zeta_A : [0, \infty) \to [0, \infty]$ is defined by

$$\zeta_A(s) = \sum_{n \in A} n^{-s}.$$

It is easy to see (and was proven by Cahen [6] in 1894; see also [2, 13]) that zeta-dimension admits the "entropy characterization"

$$\text{Dim}_\zeta(A) = \limsup_{n \to \infty} \frac{\log |A \cap \{1, \ldots, n\}|}{\log n}. \tag{1.3}$$

It is then natural to define the *lower zeta-dimension* of $A$ to be

$$\text{dim}_\zeta(A) = \liminf_{n \to \infty} \frac{\log |A \cap \{1, \ldots, n\}|}{\log n}. \tag{1.4}$$

Various properties of zeta-dimension and lower zeta-dimension, along with extensive historical references, appear in the recent paper [11], but none of this material is needed to follow our technical arguments in the present paper.

It is evident that a set $A \subseteq \mathbb{Z}^+$ satisfies the Copeland-Erdös hypothesis if and only if $\dim_\zeta(A) = 1$. The Copeland-Erdös result thus says that, for all infinite $A \subseteq \mathbb{Z}^+$ and $k \geq 2$,

$$\dim_\zeta(A) = 1 \implies \dim_{\text{FS}}(\text{CE}_k(A)) = 1. \tag{1.5}$$

Our main theorem extends (1.5) by showing that, for all infinite $A \subseteq \mathbb{Z}^+$ and $k \geq 2$,

$$\dim_{\text{FS}}(\text{CE}_k(A)) \geq \dim_\zeta(A), \tag{1.6}$$

and, dually,

$$\text{Dim}_{\text{FS}}(\text{CE}_k(A)) \geq \text{Dim}_\zeta(A). \tag{1.7}$$

Moreover, these bounds are tight in the following strong sense. Let $A \subseteq \mathbb{Z}^+$ be infinite, let $k \geq 2$, and let $\alpha = \dim_\zeta(A)$, $\beta = \text{Dim}_\zeta(A)$, $\gamma = \dim_{\text{FS}}(\text{CE}_k(A))$, $\delta = \text{Dim}_{\text{FS}}(\text{CE}_k(A))$. Then, by (1.6), (1.7), and elementary properties of these dimensions, we must have the inequalities

$$\begin{array}{ccc} \gamma \leq & \delta & \leq 1 \\ \vee\!| & \vee\!| & \\ 0 \leq \alpha \leq & \beta. & \end{array} \tag{1.8}$$

Our main theorem also shows that, for *any* $\alpha$, $\beta$, $\gamma$, $\delta$ satisfying (1.8) and any $k \geq 2$, there is an infinite set $A \subseteq \mathbb{Z}^+$ such that $\dim_\zeta(A) = \alpha$, $\text{Dim}_\zeta(A) = \beta$, $\dim_{\text{FS}}(\text{CE}_k(A)) = \gamma$, and $\text{Dim}_{\text{FS}}(\text{CE}_k(A)) = \delta$. Thus the inequalities

$$\begin{array}{ccc} \dim_{\text{FS}}(\text{CE}_k(A)) \leq & \text{Dim}_{\text{FS}}(\text{CE}_k(A)) & \leq 1 \\ \vee\!| & \vee\!| & \\ 0 \leq \quad \dim_\zeta(A) \quad \leq & \text{Dim}_\zeta(A). & \end{array} \tag{1.9}$$

are the *only* constraints that these four quantities obey in general.

The rest of this paper is organized as follows. Section 2 presents basic notation and terminology. Section 3 reviews the definitions of finite-state dimension and finite-state strong dimension and gives useful characterizations of zeta-dimension and lower zeta-dimension. Section 4 presents our main theorem.

## 2 Preliminaries

We write $\mathbb{Z}^+ = \{1, 2, \ldots\}$ for the set of positive integers. For an infinite set $A \subseteq \mathbb{Z}^+$, we often write $A = \{a_1 < a_2 < \cdots\}$ to indicate that $a_1, a_2, \ldots$ is an enumeration of $A$ in increasing numerical order. The quantifier $\exists^\infty n$ means "there exist infinitely many $n \in \mathbb{Z}^+$ such that $\ldots$", while the dual quantifier $\forall^\infty n$ means "for all but finitely many $n \in \mathbb{Z}^+$, $\ldots$".

We work in the alphabets $\Sigma_k = \{0, 1, \ldots, k-1\}$ for $k \geq 2$. The set of all (finite) *strings* over $\Sigma_k$ is $\Sigma_k^*$, and the set of all (infinite) *sequences* over $\Sigma_k$ is $\Sigma_k^\infty$. We write $\lambda$ for the empty string. Given a sequence $S \in \Sigma_k^\infty$ and integers $0 \leq i \leq j$, we write $S[i..j]$ for the string consisting of the $i$th through $j$th symbols in $S$. In particular, $S[0..n-1]$ is the string consisting of the first $n$ symbols of $S$. We write $w \sqsubseteq z$ to indicate that the string $w$ is a prefix of the string or sequence $z$.

We use the notation $\Delta(\Sigma_k)$ for the set of all probability measures on $\Sigma_k$, i.e., all functions $\pi : \Sigma_k \to [0,1]$ satisfying $\Sigma_{a \in \Sigma_k} \pi(a) = 1$. Identifying each probability measure $\pi \in \Delta(\Sigma_k)$ with the vector $(\pi(0), \ldots, \pi(k-1))$ enables us to regard $\Delta(\Sigma_k)$ as a closed simplex in the $k$-dimensional

Euclidean space $\mathbb{R}^k$. We write $\Delta_{\mathbb{Q}}(\Sigma_k)$ for the set of all rational-valued probability measures $\pi \in \Delta(\Sigma_k)$. It is often convenient to represent a positive probability measure $\pi \in \Delta_{\mathbb{Q}}(\Sigma_k)$ by a vector $\vec{a} = (a_0, \ldots, a_{k-1})$ of positive integers such that, for all $i \in \Sigma_k$, $\pi(i) = \frac{a_i}{n}$, where $n = \sum_{i=0}^{k-1} a_i$. In this case, $\vec{a}$ is called a *partition of* $n$. When $\vec{a}$ represents $\pi$ in this way, we write $\pi = \frac{\vec{a}}{n}$.

The $k$-ary *Shannon entropy* [9] of a probability measure $\pi \in \Delta(\Sigma_k)$ is

$$\mathcal{H}_k(\pi) = \mathrm{E}_\pi \log_k \frac{1}{\pi(i)} = \sum_{i=0}^{k-1} \pi(i) \log_k \frac{1}{\pi(i)},$$

where $\mathrm{E}_\pi$ denotes mathematical expectation relative to the probability measure $\pi$ and we stipulate that $0 \log_k \frac{1}{0} = 0$, so that $\mathcal{H}_k$ is continuous on the simplex $\Delta(\Sigma_k)$. The $k$-ary *Kullback-Leibler divergence* [9] between probability measures $\pi, \tau \in \Delta(\Sigma_k)$ is

$$\mathcal{D}_k(\pi \parallel \tau) = \mathrm{E}_\pi \log_k \frac{\pi(i)}{\tau(i)} = \sum_{i=0}^{k-1} \pi(i) \log_k \frac{\pi(i)}{\tau(i)}.$$

It is well-known that $\mathcal{D}_k(\pi \parallel \tau) \geq 0$, with equality if and only if $\pi = \tau$.

For $k \geq 2$ and $n \in \mathbb{Z}^+$, we write $\sigma_k(n)$ for the standard base-$k$ representation of $n$. Note that $\sigma_k(n) \in \Sigma_k^*$ and that the length of (number of symbols in) $\sigma_k(n)$ is $|\sigma_k(n)| = 1 + \lfloor \log_k n \rfloor$. Note also that, if $A = \{a_1 < a_2 < \cdots\} \subseteq \mathbb{Z}^+$ is infinite, then the base-$k$ Copeland-Erdös sequence of $A$ is

$$\mathrm{CE}_k(A) = \sigma_k(a_1)\sigma_k(a_2)\cdots \in \Sigma_k^\infty.$$

Given a set $A \subseteq \mathbb{Z}^+$ and $k, n \in \mathbb{Z}^+$, we write $A_{=n} = \{a \in A \mid |\sigma_k(a)| = n\}$ in contexts where the base $k$ is clear.

We write $\log n$ for $\log_2 n$.

## 3 The Four Dimensions

As promised in the introduction, this section gives precise definitions of finite-state dimension and finite-state strong dimension. It also gives a useful bound on the success of finite-state gamblers and useful characterizations of zeta-dimension and lower zeta-dimension.

**Definition.** A *finite-state gambler* (FSG) is a 5-tuple

$$G = (Q, \Sigma_k, \delta, \beta, q_0),$$

where $Q$ is a nonempty, finite set of *states*; $\Sigma_k = \{0, 1, \ldots, k-1\}$ is a finite alphabet ($k \geq 2$); $\delta : Q \times \Sigma_k \to Q$ is the *transition function*; $\beta : Q \to \Delta_{\mathbb{Q}}(\Sigma_k)$ is the *betting function*; and $q_0 \in Q$ is the *initial state*.

Finite-state gamblers have been investigated by Schnorr and Stimm [16], Feder [12], and others. The transition function $\delta$ is extended in the standard way to a function $\delta : Q \times \Sigma_k^* \to Q$. For $w \in \Sigma_k^*$, we use the abbreviation $\delta(w) = \delta(q_0, w)$.

**Definition.** ([10]). Let $G = (Q, \Sigma_k, \delta, \beta, q_0)$ be an FSG, and let $s \in [0, \infty)$. The *s-gale* of $G$ is the function

$$d_G^{(s)} : \Sigma_k^* \to [0, \infty)$$

defined by the recursion
$$d_G^{(s)}(\lambda) = 1,$$

$$d_G^{(s)}(wa) = k^s d_G^{(s)}(w)\beta(\delta(w))(a) \quad (3.1)$$

for all $w \in \Sigma_k^*$ and $a \in \Sigma_k$.

Intuitively, $d_G^{(s)}(w)$ is the amount of money that the gambler $G$ has after betting on the successive symbols in the string $w$. The parameter $s$ controls the payoffs via equation (3.1). If $s = 1$, then the payoffs are fair in the sense that the conditional expected value of $d_G^{(1)}(wa)$, given that $w$ has occurred and the symbols $a \in \Sigma_k$ are all equally likely to follow $w$, is precisely $d_G^{(1)}(w)$. If $s < 1$, then the payoffs are unfair.

We repeatedly use the obvious fact that $d_G^{(s)}(w) \le k^{s|w|}$ holds for all $s$ and $w$.

**Definition.** Let $G = (Q, \Sigma_k, \delta, \beta, q_0)$ be an FSG, let $s \in [0, \infty)$, and let $S \in \Sigma_k^\infty$.

1. $G$ *s-succeeds* on $S$ if
$$\limsup_{n \to \infty} d_G^{(s)}(S[0..n-1]) = \infty.$$

2. $G$ *strongly s-succeeds* on $S$ if
$$\liminf_{n \to \infty} d_G^{(s)}(S[0..n-1]) = \infty.$$

**Definition.** Let $S \in \Sigma_k^\infty$.

1. [10]. The *finite-state dimension* of $S$ is
$$\dim_{\mathrm{FS}}(S) = \inf \{s \mid \text{there is an FSG that } s\text{-succeeds on } S\}.$$

2. [3] The *finite-state strong dimension* of $S$ is
$$\mathrm{Dim}_{\mathrm{FS}}(S) = \inf \{s \mid \text{there is an FSG that strongly } s\text{-succeeds on } S\}.$$

It is easy to verify that $0 \le \dim_{\mathrm{FS}}(S) \le \mathrm{Dim}_{\mathrm{FS}}(S) \le 1$ for all $S \in \Sigma_k^\infty$. More properties of these finite-state dimensions, including their relationships to classical Hausdorff and packing dimensions, respectively, may be found in [10, 3].

It is useful to have a measure of the size of a finite-state gambler. This size depends on the alphabet size, the number of states, and the least common denominator of the values of the betting function in the following way.

**Definition.** The *size* of an FSG $G = (Q, \Sigma_k, \delta, \beta, q_0)$ is
$$\mathrm{size}(G) = (k+l)|Q|,$$
where $l = \min \{l \in \mathbb{Z}^+ \mid (\forall q \in Q)(\forall i \in \Sigma_k) l\beta(q)(i) \in \mathbb{Z}\}$.

**Observation 3.1.** *For each $k \ge 2$ and $t \in \mathbb{Z}^+$, there are, up to renaming of states, fewer than $t^2(2t)^t$ finite-state gamblers $G$ with $\mathrm{size}(G) \le t$.*

*Proof.* Given $k, l, m \in \mathbb{Z}^+$ with $k \geq 2$, let $\mathcal{G}_{k,l,m}$ be the set of all FSGs $G = (\Sigma_m, \Sigma_k, \delta, \beta, q_0)$ satisfying $l\beta(q)(i) \in \mathbb{Z}$ for all $q \in \Sigma_m$ and $i \in \Sigma_k$. Equivalently, $\mathcal{G}_{k,l,m}$ is the set of all FSGs $G = (Q, \Sigma_k, \delta, \beta, q_0)$ such that $Q = \{0, \ldots, m-1\}$ and $\beta : Q \to \Delta_{\mathbb{Q}_l}(\Sigma_k)$, where

$$\Delta_{\mathbb{Q}_l}(\Sigma_k) = \{\pi \in \Delta_{\mathbb{Q}}(\Sigma_k) \mid (\forall i \in \Sigma_k) l\pi(i) \in \mathbb{Z}\}.$$

Since $|\Delta_{\mathbb{Q}_l}(\Sigma_k)| = \binom{k+l-1}{k-1}$, it is easy to see that

$$|\mathcal{G}_{k,l,m}| = m^{km+1} \binom{k+l-1}{k-1}^m. \tag{3.2}$$

Now fix $k \geq 2$ and $t \in \mathbb{Z}^+$, and let $\mathcal{G}_t$ be the set of all FSGs $G = (\Sigma_m, \Sigma_k, \delta, \beta, q_0)$ with $\text{size}(G) \leq t$. Our objective is to show that $|\mathcal{G}_t| < t^2(2t)^t$. For each $1 \leq j \leq t$, there are at most $j$ pairs $(l, m)$ such that $(k+l)m = j$, and, for each of these pairs $(l, m)$, (3.2) tells us that $|\mathcal{G}_{k,l,m}| < (2j)^j$, so

$$|\mathcal{G}_t| < \sum_{j=1}^{t} j(2j)^j < t^2(2t)^t.$$

$\square$

In general, an *s-gale* is a function $d : \Sigma_k^* \to [0, \infty)$ satisfying

$$d(w) = k^{-s} \sum_{a=0}^{k-1} d(wa)$$

for all $w \in \Sigma_k^*$ [15]. It is clear that $d_G^{(s)}$ is an $s$-gale for every FSG $G$ and every $s \in [0, \infty)$. The case $k = 2$ of the following lemma was proven in [15]. The extension to arbitrary $k \geq 2$ is routine.

**Lemma 3.2.** ([15]). *If $s \in [0, 1]$ and $d$ is an $s$-gale, then, for all $w \in \Sigma_k^*$, $j \in \mathbb{N}$, and $0 < \alpha \in \mathbb{R}$, there are fewer than $\frac{k^{sj}}{\alpha}$ strings $u \in \Sigma_k^*$ of length $j$ for which $d(u) > \alpha$.*

The following lemma will be useful in proving our main theorem.

**Lemma 3.3.** *For each $s, \alpha \in (0, \infty)$ and $k, n, t \in \mathbb{Z}^+$ with $k \geq 2$, there are fewer than*

$$\frac{k^{2s} n^s t^2 (2t)^t}{\alpha(k^s - 1)}$$

*integers $m \in \{1, \ldots, n\}$ for which*

$$\max_{\text{size}(G) \leq t} d_G^{(s)}(\sigma_k(m)) \geq \alpha,$$

*where the maximum is taken over all FSGs $G = (Q, \Sigma_k, \delta, \beta, q_0)$ with $\text{size}(G) \leq t$.*

*Proof.* Let $s, \alpha, k, n$, and $t$ be as given, and let $\mathcal{G}_t$ be the set of all FSGs $G = (\Sigma_m, \Sigma_k, \delta, \beta, q_0)$ with $\text{size}(G) \leq t$. For each $j \in \mathbb{Z}^+$ and $G \in \mathcal{G}_t$, Lemma 3.2 tells us that there are fewer than $\frac{k^{sj}}{\alpha}$ strings $u \in \Sigma_k^*$ of length $j$ for which $d_G^{(s)}(u) \geq \alpha$. It follows by Observation 3.1 that, for each $j \in \mathbb{Z}^+$, there are fewer than $t^2(2t)^t \frac{k^{sj}}{\alpha}$ strings $u \in \Sigma_k^*$ of length $j$ for which

$$\max_{G \in \mathcal{G}_t} d_G^{(s)}(u) \geq \alpha$$

holds. Since

$$\sum_{j=1}^{|\sigma_k(n)|} t^2(2t)^t \frac{k^{sj}}{\alpha} = \frac{t^2(2t)^t}{\alpha} \sum_{j=1}^{1+\lfloor \log_k n \rfloor} k^{sj} \leq \frac{k^{2s} n^s t^2 (2t)^t}{\alpha(k^s - 1)},$$

the lemma follows. $\square$

The zeta-dimension $\text{Dim}_\zeta(A)$ and lower zeta-dimension $\dim_\zeta(A)$ of a set $A$ of positive integers were defined in the introduction. The following lemma gives useful characterizations of these quantities in terms of the increasing enumeration of $A$.

**Lemma 3.4.** *Let $A = \{a_1 < a_2 < \cdots\}$ be an infinite set of positive integers.*

1. $\dim_\zeta(A) = \inf \left\{ t \geq 0 \mid (\exists^\infty n) a_n^t > n \right\} = \inf \left\{ t \geq 0 \mid (\exists^\infty n) a_n^t \geq n \right\}$
   $= \sup \left\{ t \geq 0 \mid (\forall^\infty n) a_n^t < n \right\} = \sup \left\{ t \geq 0 \mid (\forall^\infty n) a_n^t \leq n \right\}.$

2. $\text{Dim}_\zeta(A) = \inf \left\{ t \geq 0 \mid (\forall^\infty n) a_n^t > n \right\} = \inf \left\{ t \geq 0 \mid (\forall^\infty n) a_n^t \geq n \right\}$
   $= \sup \left\{ t \geq 0 \mid (\exists^\infty n) a_n^t < n \right\} = \sup \left\{ t \geq 0 \mid (\exists^\infty n) a_n^t \leq n \right\}.$

*Proof.* Let $A$ be as given. For each $R \in \{<, \leq, >, \geq\}$, define the sets

$$I_R = \left\{ t \geq 0 \mid (\exists_n^\infty) a_n^t \, R \, n \right\},$$

$$J_R = \left\{ t \geq 0 \mid (\forall^\infty n) a_n^t \, R \, n \right\}.$$

Our task is then to prove that

$$\dim_\zeta(A) = \inf I_> = \inf I_\geq = \sup J_< = \sup J_\leq \tag{3.3}$$

and

$$\text{Dim}_\zeta(A) = \inf J_> = \inf J_\geq = \sup I_< = \sup I_\leq. \tag{3.4}$$

Note that each of the pairs $(J_<, I_\geq)$, $(J_\leq, I_>)$, $(I_<, J_\geq)$, $(I_\leq, J_>)$ partitions $[0, \infty)$ into two nonempty subsets with every element of the left component less than every element of the right component, the left components satisfying

$$0 \in J_< \subseteq J_\leq \cap I_< \subseteq J_\leq \cup I_< \subseteq I_\leq,$$

and the right components satisfying

$$(1, \infty) \subseteq J_> \subseteq J_\geq \cap I_> \subseteq J_\geq \cup I_> \subseteq I_\geq.$$

It follows immediately from this that

$$\sup J_< = \inf I_\geq \leq \sup J_\leq = \inf I_>$$

and

$$\sup I_< = \inf J_\geq \leq \sup I_\leq = \inf J_>.$$

Hence, to prove (3.3) and (3.4), it suffices to show that

$$\inf I_> \leq \dim_\zeta(A) \leq \inf I_\geq \tag{3.5}$$

$$\inf J_> \leq \mathrm{Dim}_\zeta(A) \leq \inf J_\geq. \tag{3.6}$$

To see that $\inf I_> \leq \dim_\zeta(A)$, let $t > \dim_\zeta(A)$. Fix $t'$ with $t > t' > \dim_\zeta(A)$. Then, by the definition of $\dim_\zeta(A)$, there exist infinitely many $n \in \mathbb{Z}^+$ such that

$$|A \cap \{1, \ldots, n\}| < n^{t'}. \tag{3.7}$$

If $n$ satisfies (3.7) and is large enough that $n^t \geq n^{t'} + 1$, fix $k$ such that $a_k \leq n < a_{k+1}$. Then we have

$$a_{k+1}^t > n^t \geq n^{t'} + 1 > |A \cap \{1, \ldots, n\}| + 1 = k + 1.$$

It follows that there exist infinitely many $k$ such that $a_k^t > k$, i.e., that $t \in I_>$, whence $\inf I_> \leq t$. Since this holds for all $t > \dim_\zeta(A)$, it follows that $\inf I_> \leq \dim_\zeta(A)$.

To see that $\dim_\zeta(A) \leq \inf I_\geq$, let $t > \inf I_\geq$. Then there exist infinitely many $n \in \mathbb{Z}^+$ such that $a_n^t \geq n$. For each of these $n$, we have

$$|A \cap \{1, \ldots, a_n\}| = n \leq a_n^t,$$

so there exist infinitely many $m \in \mathbb{Z}^+$ such that

$$|A \cap \{1, \ldots, , m\}| \leq m^t.$$

This implies that

$$\dim_\zeta(A) = \liminf_{m \to \infty} \frac{\log|A \cap \{1, \ldots, m\}|}{\log m} \leq t.$$

Since this holds for all $t > \inf I_\geq$, it follows that $\dim_\zeta(A) \leq \inf I_\geq$. This completes the proof that (3.5) holds.

The proof that (3.6) holds is similar. $\square$

## 4  Main Theorem

The proof of our main theorem uses the following combinatorial lemma.

**Lemma 4.1.** *For every $n \geq k \geq 2$ and every partition $\vec{a} = (a_0, \ldots, a_{k-1})$ of $n$, there are more than*

$$k^{n \mathcal{H}_k(\frac{\vec{a}}{n}) - (k+1)\log_k n}$$

*integers $m$ with $|\sigma_k(m)| = n$ and $\#(i, \sigma_k(m)) = a_i$ for each $i \in \Sigma_k$.*

*Proof.* Let $n \geq k \geq 2$, and let $\vec{a} = (a_0, \ldots, a_{k-1})$ be a partition of $n$. Define the sets

$$B = \{u \in \Sigma_k^n \mid (\forall i \in \Sigma_k) \#(i, u) = a_i\},$$

$$C = \{m \in \mathbb{Z}^+ \mid \sigma_k(m) \in B\}.$$

Define an equivalence relation $\sim$ on $B$ by

$$u \sim v \iff (\exists x, y \in \Sigma_k^*)[u = xy \text{ and } v = yx].$$

Then each $\sim$-equivalence class has at most $n$ elements and contains $\sigma_k(m)$ for at least one $m \in C$, so
$$|C| \geq \frac{1}{n}|B|.$$

Using multinomial coefficients and the well-known estimate $e(\frac{t}{e})^t < t! < et(\frac{t}{e})^t$, valid for all $t \in \mathbb{Z}^+$, we have
$$|B| = \binom{n}{a_0, \ldots, a_{k-1}} = \frac{n!}{\prod_{i=0}^{k-1} a_i!} > \frac{1}{e^{k-1}\prod_{i=0}^{k-1} a_i} \prod_{i=0}^{k-1}\left(\frac{n}{a_i}\right)^{a_i}.$$

Since the geometric mean is bounded by the arithmetic mean,
$$\prod_{i=0}^{k-1} a_i \leq \left(\frac{1}{k}\sum_{i=0}^{k-1} a_i\right)^k = \left(\frac{n}{k}\right)^k.$$

Putting this all together, we have
$$|C| > \frac{k^k}{e^{k-1}n^{k+1}} \prod_{i=0}^{k-1}\left(\frac{n}{a_i}\right)^{a_i} \geq \frac{1}{n^{k+1}} \prod_{i=0}^{k-1}\left(\frac{n}{a_i}\right)^{a_i},$$

whence
$$\log_k|C| > \left(\log_k \prod_{i=0}^{k-1}\left(\frac{n}{a_i}\right)^{a_i}\right) - (k+1)\log_k n$$
$$= n\mathcal{H}_k\left(\frac{\vec{a}}{n}\right) - (k+1)\log_k n.$$

$\square$

We now have all the machinery that we need to prove the main result of this paper.

**Theorem 4.2.** (main theorem). *Let $k \geq 2$.*

1. *For every infinite set $A \subseteq \mathbb{Z}^+$,*
$$\dim_{\text{FS}}(\text{CE}_k(A)) \geq \dim_\zeta(A) \tag{4.1}$$
*and*
$$\text{Dim}_{\text{FS}}(\text{CE}_k(A)) \geq \text{Dim}_\zeta(A). \tag{4.2}$$

2. *For any four real numbers $\alpha, \beta, \gamma, \delta$ satisfying the inequalities*
$$\begin{array}{c} \gamma \leq \delta \leq 1 \\ \text{VI} \quad \text{VI} \\ 0 \leq \alpha \leq \beta, \end{array} \tag{4.3}$$
*there exists an infinite set $A \subseteq \mathbb{Z}^+$ such that $\dim_\zeta(A) = \alpha$, $\text{Dim}_\zeta(A) = \beta$, $\dim_{\text{FS}}(\text{CE}_k(A)) = \gamma$, and $\text{Dim}_{\text{FS}}(\text{CE}_k(A)) = \delta$.*

*Proof.* To prove part 1, let $A = \{a_1 < a_2 < \cdots\} \subseteq \mathbb{Z}^+$ be infinite. Fix $0 < s < t < 1$, let

$$J_t = \left\{n \in \mathbb{Z}^+ \mid a_n^t < n\right\},$$

and let $G = (Q, \Sigma_k, \delta, \beta, q_0)$ be an FSG. Let $n \in \mathbb{Z}^+$, and consider the quantity $d_G^{(s)}(w_n)$, where

$$w_n = \sigma_k(a_1) \cdots \sigma_k(a_n).$$

There exist states $q_1, \ldots, q_n \in Q$ such that

$$d_G^{(s)}(w_n) = \prod_{i=1}^{n} d_{G_{q_i}}^{(s)}(\sigma_k(a_i)),$$

where $G_{q_i} = (Q, \Sigma_k, \delta, \beta, q_i)$. Let $B = \left\{1 \leq i \leq n \mid d_{G_{q_i}}^{(s)}(\sigma_k(a_i)) \geq \frac{1}{k}\right\}$, and let $B^c = \{1, \ldots, n\} - B$. Then

$$d_G^{(s)}(w_n) = \left(\prod_{i \in B} d_{G_{q_i}}^{(s)}(\sigma_k(a_i))\right) \left(\prod_{i \in B^c} d_{G_{q_i}}^{(s)}(\sigma_k(a_i))\right). \tag{4.4}$$

By our choice of $B$,

$$\prod_{i \in B^c} d_{G_{q_i}}^{(s)}(\sigma_k(a_i)) \leq k^{|B|-n}. \tag{4.5}$$

By Lemma 3.3,

$$|B| \leq \frac{ck^{2s+1}a_n^s}{k^s - 1}, \tag{4.6}$$

where $c = \text{size}(G)^2 (2\text{size}(G))^{\text{size}(G)}$. Since $d_{G_{q_i}}^{(s)}(u) \leq k^{s|u|}$ must hold in all cases, it follows that

$$\prod_{i \in B} d_{G_{q_i}}^{(s)}(\sigma_k(a_i)) \leq k^{s|B||\sigma_k(a_n)|} \leq k^{s|B|(1+\log_k a_n)}. \tag{4.7}$$

By (4.4), (4.5), (4.6), and (4.7), we have

$$\log_k d_G^{(s)}(w_n) \leq \tau(1 + s + s \log_k a_n) a_n^s - n, \tag{4.8}$$

where $\tau = \frac{ck^{2s+1}}{k^s - 1}$. If $n$ is sufficiently large, and if $n + 1 \in J_t$, then (4.8) implies that

$$\log_k d_G^{(s)}(w_n) \leq \tau(1 + s + s \log_k a_n) a_n^s - 2(n+1)^{\frac{s+t}{2t}}$$
$$\leq \tau(1 + s + s \log_k a_n) a_n^s - 2a_{n+1}^{\frac{s+t}{2}}$$
$$\leq \tau(1 + s + s \log_k a_n) a_n^s - a_n^{\frac{s+t}{2}} - s(1 + \log_k a_{n+1})$$
$$\leq -s(1 + \log_k a_{n+1})$$
$$\leq -s|\sigma_k(a_{n+1})|.$$

We have now shown that

$$d_G^{(s)}(w_n) \leq k^{-s|\sigma_k(a_{n+1})|} \tag{4.9}$$

holds for all sufficiently large $n$ with $n + 1 \in J_t$.

To prove (4.1), let $s < t < \dim_\zeta(A)$. It suffices to show that $\dim_{\mathrm{FS}}(\mathrm{CE}_k(A)) \geq s$. Since $t < \dim_\zeta(A)$, Lemma 3.4 tells us that the set $J_t$ is cofinite. Hence, for every sufficiently long prefix $w \sqsubseteq \mathrm{CE}_k(A)$, there exist $n$ and $u \sqsubseteq \sigma_k(a_{n+1})$ such that $w = w_n u$ and (4.9) holds, whence

$$d_G^{(s)}(w) \leq k^{-s|\sigma_k(a_{n+1})|} k^{s|u|} \leq 1.$$

This shows that $G$ does not $s$-succeed on $\mathrm{CE}_k(A)$, whence $\dim_{\mathrm{FS}}(\mathrm{CE}_k(A)) \geq s$.

To prove (4.2), let $s < t < \mathrm{Dim}_\zeta(A)$. It suffices to show that $\mathrm{Dim}_{\mathrm{FS}}(\mathrm{CE}_k(A)) \geq s$. Since $t < \mathrm{Dim}_\zeta(A)$, Lemma 3.4 tells us that the set $J_t$ is infinite. For the infinitely many $n$ for which $n + 1 \in J_t$ and (4.9) holds, we then have $d_G^{(s)}(w_n) < 1$. This shows that $G$ does not strongly $s$-succeed on $\mathrm{CE}_k(A)$, whence $\mathrm{Dim}_{\mathrm{FS}}(\mathrm{CE}_k(A)) \geq s$.

To prove part 2 of the theorem, let $\alpha$, $\beta$, $\gamma$, and $\delta$ be real numbers satisfying (4.3). We will explicitly construct an infinite set $A \subseteq \mathbb{Z}^+$ with the indicated dimensions. Intuitively, the values of $\dim_\zeta(A)$ and $\mathrm{Dim}_\zeta(A)$ will be achieved by controlling the density of $A$; the upper bounds on $\dim_{\mathrm{FS}}(\mathrm{CE}_k(A))$ and $\mathrm{Dim}_{\mathrm{FS}}(\mathrm{CE}_k(A))$ will be achieved by constructing $A$ from integers whose base-$k$ expansions have controlled frequencies of digits (such integers being abundant by Lemma 4.1); and the lower bounds on $\dim_{\mathrm{FS}}(\mathrm{CE}_k(A))$ and $\mathrm{Dim}_{\mathrm{FS}}(\mathrm{CE}_k(A))$ will be achieved by avoiding use of the very few (by Lemma 3.3) integers on whose base-$k$ expansions a finite-state gambler can win.

We first define some useful probability measures on $\Sigma_k$, all expressed as vectors. Let $\vec{\mu} = (\frac{1}{k}, \ldots, \frac{1}{k}) \in \Delta(\Sigma_k)$ be the uniform probability measure, and let $\vec{\nu} = (1, 0, \ldots, 0) \in \Delta(\Sigma_k)$ be the degenerate probability measure that concentrates all probability on 0. Define the function $g : [0, 1] \to \Delta(\Sigma_k)$ by

$$g(r) = r\vec{\mu} + (1 - r)\vec{\nu}.$$

Then $g$ defines a line segment from a corner $g(0) = \vec{\nu}$ to the centroid $g(1) = \vec{\mu}$ of the simplex $\Delta(\Sigma_k)$. Also, $\mathcal{H}_k \circ g : [0, 1] \to [0, 1]$ is strictly increasing and continuous, with $\mathcal{H}_k(g(0)) = 0$ and $\mathcal{H}_k(g(1)) = 1$. Let $r_\gamma = (\mathcal{H}_k \circ g)^{-1}(\gamma)$, $r_\delta = (\mathcal{H}_k \circ g)^{-1}(\delta)$, $\vec{\pi} = g(r_\gamma)$, and $\vec{\tau} = g(r_\delta)$, so that

$$\mathcal{H}_k(\vec{\pi}) = \gamma, \mathcal{H}_k(\vec{\tau}) = \delta.$$

Then let $\vec{\pi}^{(k)}, \vec{\pi}^{(k+1)}, \vec{\pi}^{(k+2)}, \ldots$ and $\vec{\tau}^{(k)}, \vec{\tau}^{(k+1)}, \vec{\tau}^{(k+2)}, \ldots$ be sequences in $\Delta_\mathbb{Q}(\Sigma_k)$ with the following properties.

(i) For each $n \geq k$, $n\vec{\pi}^{(n)}$ and $n\vec{\tau}^{(n)}$ are partitions of $n$, with each $n\pi_i^{(n)} \geq \sqrt{n}$ and $n\tau_i^{(n)} \geq \sqrt{n}$ for $n \geq k^2$.

(ii) $\lim_{n \to \infty} \vec{\pi}^{(n)} = \vec{\pi}$ and $\lim_{n \to \infty} \vec{\tau}^{(n)} = \vec{\tau}$.

Note that (i) ensures that

$$\mathcal{H}_k(\vec{\pi}^{(n)}) \geq \frac{k-1}{2\sqrt{n}} \log_k n, \quad \mathcal{H}_k(\vec{\tau}^{(n)}) \geq \frac{k-1}{2\sqrt{n}} \log_k n \tag{4.10}$$

hold for all $n \geq k^2$.

For each $u \in \Sigma_k^*$ and $s \in [0, \infty)$, let $\mathcal{G}_u$ be the set of all FSGs $G$ with $\mathrm{size}(G) \leq \log_k \log_k |u|$, and let

$$d_{\max}^{(s)}(u) = \max_{G \in \mathcal{G}_u} d_G^{(s)}(u).$$

Define the sets

$$U = \left\{ a \geq k^{k-1} \;\middle|\; d_{\max}^{(\mathcal{H}_k(\vec{\pi}^{(|\sigma_k(a)|)}))}(\sigma_k(a)) > |\sigma_k(a)|^{k+2} \right\},$$
$$V = \left\{ a \geq k^{k-1} \;\middle|\; d_{\max}^{(\mathcal{H}_k(\vec{\tau}^{(|\sigma_k(a)|)}))}(\sigma_k(a)) > |\sigma_k(a)|^{k+2} \right\},$$
$$C = \left\{ a \geq k^{k-1} \;\middle|\; (\forall i \in \Sigma_k) \#(i, \sigma_k(a)) = |\sigma_k(a)| \pi_i^{(|\sigma_k(a)|)} \right\},$$
$$D = \left\{ a \geq k^{k-1} \;\middle|\; (\forall i \in \Sigma_k) \#(i, \sigma_k(a)) = |\sigma_k(a)| \tau_i^{(|\sigma_k(a)|)} \right\},$$
$$C' = C - U,$$
$$D' = D - V.$$

Then, for all $n \geq k$, we have

$$|U_{=n}| = \left\{ a \in \mathbb{Z}_{=n}^+ \;\middle|\; d_{\max}^{(\mathcal{H}_k(\vec{\pi}^{(n)}))}(\sigma_k(a)) > n^{k+2} \right\},$$

so Lemma 3.3 tells us that

$$|U_{=n}| < \frac{k^{2\mathcal{H}_k(\vec{\pi}^{(n)}) + n\mathcal{H}_k(\vec{\pi}^{(n)})} t^2 (2t)^t}{n^{k+2} (k^{\mathcal{H}_k(\vec{\pi}^{(n)})} - 1)}$$

for all $n \geq k$, where $t = \log_k \log_k n$. It follows easily from this that

$$|U_{=n}| = o(k^{n\mathcal{H}_k(\vec{\pi}^{(n)}) - (k+1)\log_k n}) \tag{4.11}$$

as $n \to \infty$. By Lemma 4.1, we have

$$|C_{=n}| \geq k^{n\mathcal{H}_k(\vec{\pi}^{(n)}) - (k+1)\log_k n}. \tag{4.12}$$

(By (4.10), this is positive for all sufficiently large $n$.) Putting (4.11) and (4.12) together with our choice of the $\vec{\pi}^{(n)}$ gives us

$$|C'_{=n}| \geq \max\{1, k^{(\alpha - o(1))n}\} \tag{4.13}$$

as $n \to \infty$. A similar argument shows that

$$|D'_{=n}| \geq \max\{1, k^{(\beta - o(1))n}\} \tag{4.14}$$

as $n \to \infty$. It follows that we can fix sets $C'' \subseteq C'$ and $D'' \subseteq D'$ such that

$$\max\{1, k^{(\alpha - o(1))n}\} \leq |C''_{=n}| \leq k^{(\alpha + o(1))n} \tag{4.15}$$

and

$$\max\{1, k^{(\beta - o(1))n}\} \leq |D''_{=n}| \leq k^{(\beta + o(1))n} \tag{4.16}$$

as $n \to \infty$.

Now define $T : \mathbb{Z}^+ \to \mathbb{Z}^+$ by the recursion

$$T(1) = k, T(l+1) = k^{T(l)},$$

so that $T(l)$ is an "exponential tower" $k^{k^{\cdot^{\cdot^{\cdot^k}}}}$ of height $l$. For each $n \geq k$, let $T^{-1}(n)$ be the unique $l$ such that $T(l) \leq n < T(l+1)$. Let

$$C^* = \bigcup_{T^{-1}(n) \text{ even}} C''_{=n}, \quad D^* = \bigcup_{T^{-1}(n) \text{ odd}} D''_{=n},$$

and let

$$A = C^* \cup D^*.$$

This is our set $A$.

We now note the following.

1. By (4.15),

$$|A \cap \{1, \ldots, k^{T(2l+1)-2}\}|$$
$$= \sum_{n=1}^{T(2l)-1} |A_{=n}| + \sum_{n=T(2l)}^{T(2l+1)-1} |A_{=n}|$$
$$\leq \sum_{n=0}^{T(2l)-1} k^n + \sum_{n=T(2l)}^{T(2l+1)-1} k^{(\alpha+o(1))n}$$
$$\leq k^{T(2l)} + k^{(\alpha+o(1))T(2l+1)}$$
$$= k^{(\alpha+o(1))T(2l+1)}$$

as $l \to \infty$, so (1.4) tells us that

$$\dim_\zeta(A) \leq \liminf_{l \to \infty} \frac{\log_k |A \cap \{1, \ldots, k^{T(2l+1)-2}\}|}{\log_k k^{T(2l+1)-2}}$$
$$\leq \liminf_{l \to \infty} \frac{(\alpha + o(1))T(2l+1)}{T(2l+1) - 2} = \alpha.$$

2. By (4.15), (4.16), and the fact that $\alpha \leq \beta$,

$$|A \cap \{1, \ldots, m\}| \geq \sum_{n=1}^{|\sigma_k(m)|-1} |A_{=n}|$$
$$\geq \sum_{n=1}^{|\sigma_k(m)|-1} k^{(\alpha-o(1))n}$$
$$= k^{(\alpha-o(1))|\sigma_k(m)|}$$
$$= m^{\alpha-o(1)}$$

as $m \to \infty$, so (1.4) tells us that $\dim_\zeta(A) \geq \alpha$.

3. By (4.15), (4.16), and the fact that $\alpha \leq \beta$,

$$|A \cap \{1, \ldots, m\}| \leq \sum_{n=1}^{|\sigma_k(m)|} |A_{=n}|$$

$$\leq \sum_{n=1}^{|\sigma_k(m)|} k^{(\beta+o(1))n}$$

$$= k^{(\beta+o(1))|\sigma_k(m)|}$$

$$= m^{\beta+o(1)}$$

as $m \to \infty$, so (1.3) tells us that $\mathrm{Dim}_\zeta(A) \leq \beta$.

4. By (1.3) and (4.16),

$$\mathrm{Dim}_\zeta(A) \geq \limsup_{n \to \infty} \frac{\log_k |A_{=n}|}{\log_k(k^n - 1)}$$

$$\geq \limsup_{n \to \infty} \frac{\log_k k^{(\beta-o(1))n}}{\log_k(k^n - 1)} = \beta.$$

These four things together show that $\dim_\zeta(A) = \alpha$ and $\mathrm{Dim}_\zeta(A) = \beta$.

Our next objective is to prove that $\dim_{\mathrm{FS}}(\mathrm{CE}_k(A)) \geq \gamma$ and $\mathrm{Dim}_{\mathrm{FS}}(\mathrm{CE}_k(A)) \geq \delta$. For this, let $G = (Q, \Sigma_k, \delta, \beta, q_0)$ be an FSG, and let $s \in [0, \infty)$. It suffices to prove that

$$s < \gamma \Rightarrow G \text{ does not } s\text{-succeed on } \mathrm{CE}_k(A) \tag{4.17}$$

and

$$s < \delta \Rightarrow G \text{ does not strongly } s\text{-succeed on } \mathrm{CE}_k(A). \tag{4.18}$$

Write $A = \{a_1 < a_2 < \cdots\}$, so that

$$\mathrm{CE}_k(A) = \sigma_k(a_1)\sigma_k(a_2)\sigma_k(a_3)\cdots.$$

There is a sequence $q_1, q_2, q_3, \ldots$ of states $q_i \in Q$ such that, for any $m \geq 0$ and any proper prefix $u \sqsubsetneq \sigma_k(a_{m+1})$,

$$d_G^{(s)}(\sigma_k(a_1) \cdots \sigma_k(a_m)u) = \left(\prod_{i=0}^{m-1} d_{G_{q_i}}^{(s)}(\sigma_k(a_{i+1}))\right) d_{G_{q_m}}^{(s)}(u), \tag{4.19}$$

where $G_q = (Q, \Sigma_k, \delta, \beta, q)$. Let $c = \mathrm{size}(G)$. Note that, for all $q \in Q$, $\mathrm{size}(G_q) = c$, so

$$a \geq k^{k^{k^c}} \Rightarrow c \leq \log_k \log_k \log_k a \leq \log_k \log_k |\sigma_k(a)|$$

$$\Rightarrow G_q \in \mathcal{G}_{\sigma_k(a)}.$$

Since $C^* \cap U = \varnothing$, it follows that, for all $q \in Q$,

$$k^{k^{k^c}} \leq a \in C^*_{=n} \Rightarrow d_{G_q}^{(\mathcal{H}_k(\vec{\pi}^{(n)}))}(\sigma_k(a)) \leq n^{k+2}.$$

Using the identity $d_{G_q}^{(s)}(x) = k^{(s-s')|x|} d_{G_q}^{(s')}(x)$ and the facts that $\mathcal{H}_k(\vec{\pi}^{(n)}) = \gamma + o(1)$ and $n^{k+2} = k^{o(n)}$ as $n \to \infty$, we then have, for all $q \in Q$,

$$a \in C_{=n}^* \Rightarrow d_{G_q}^{(s)}(\sigma_k(a)) \leq k^{(s-\gamma+o(1))n} \tag{4.20}$$

as $n \to \infty$. A similar argument shows that, for all $q \in Q$,

$$a \in D_{=n}^* \Rightarrow d_{G_q}^{(s)}(\sigma_k(a)) \leq k^{(s-\delta+o(1))n} \tag{4.21}$$

as $n \to \infty$.

To verify (4.17), assume that $s < \gamma$. Then, since $\gamma \leq \delta$, (4.20) and (4.21) tell us that

$$d_{G_{q_i}}^{(s)}(\sigma_k(a_{i+1})) \leq k^{(s-\gamma+o(1))|\sigma_k(a_{i+1})|}$$

as $i \to \infty$. It follows by (4.19) that, for any prefix $w \sqsubseteq \mathrm{CE}_k(A)$, if we write $w = \sigma_k(a_1) \cdots \sigma_k(a_m) u$, where $u \sqsubsetneq \sigma_k(a_{m+1})$, then $|u| = o(|w|)$ as $|w| \to \infty$, so

$$d_G^{(s)}(w) \leq \left( \prod_{i=0}^{m-1} k^{(s-\gamma+o(1))|\sigma_k(a_{i+1})|} \right) k^{s|u|}$$
$$= k^{(s-\gamma+o(1))(|w|-|u|)+s|u|}$$
$$= k^{(s-\gamma+o(1))|w|}$$

as $|w| \to \infty$. Since $s < \gamma$, it follows that

$$\limsup_{n \to \infty} d_G^{(s)}(\mathrm{CE}_k(A)[0..n-1]) = 0,$$

affirming (4.17).

To verify (4.18), assume that $s < \delta$. For each $l \in \mathbb{Z}^+$, let

$$v_l = \sigma_k(a_{i_l}) \sigma_k(a_{i_l+1}) \cdots \sigma_k(a_{i_{l+1}-1}),$$

where $i_l$ is the least $i$ such that $|\sigma_k(a_i)| = T(l)$, and let

$$w_l = v_1 v_2 \cdots v_{l-1},$$

noting that each $w_l \sqsubseteq \mathrm{CE}_k(A)$. Then $|w_l| = o(|v_l|)$ as $l \to \infty$, so

$$d_G^{(s)}(w_{2l}) = d_G^{(s)}(w_{2l-1}) \prod_{i=i_{2l-1}}^{i_{2l}-1} d_{G_{q_{i-1}}}^{(s)}(\sigma_k(a_i))$$
$$\leq k^{s|w_{2l-1}|} \prod_{i=i_{2l-1}}^{i_{2l}-1} k^{(s-\delta+o(1))|\sigma_k(a_i)|}$$
$$= k^{s|w_{2l-1}|+(s-\delta+o(1))|v_{2l-1}|}$$
$$= k^{(s-\delta+o(1))|v_{2l-1}|}$$

as $l \to \infty$. Since $s < \delta$, this affirms (4.18) and concludes the proof that $\dim_{\mathrm{FS}}(\mathrm{CE}_k(A)) \geq \gamma$ and $\mathrm{Dim}_{\mathrm{FS}}(\mathrm{CE}_k(A)) \geq \delta$.

All that remains is to prove that $\dim_{\mathrm{FS}}(\mathrm{CE}_k(A)) \leq \gamma$ and $\mathrm{Dim}_{\mathrm{FS}}(\mathrm{CE}_k(A)) \leq \delta$. For each rational $r \in \mathbb{Q} \cap [0,1]$, let $G_r$ be the 1-state FSG whose bets are given by $g(r)$, where $g : [0,1] \to \Delta(\Sigma_k)$ is the function defined earlier in this proof. That is, for all $s \in [0, \infty)$, $w \in \Sigma_k^*$, and $a \in \Sigma_k$, we have
$$d_{G_r}^{(s)}(wa) = k^s g(r)(a) d_{G_r}^{(s)}(w).$$

If we write $\theta_w(a) = \frac{\#(a,w)}{|w|}$ for all $w \in \Sigma_k^+$ and $a \in \Sigma_k$, then this implies that, for all $w \in \Sigma_k^+$,
$$d_{G_r}^{(s)}(w) = k^{s|w|} \prod_{a \in \Sigma_k} g(r)(a)^{\#(a,w)},$$

whence
$$\log_k d_{G_r}^{(s)}(w) = s|w| + \sum_{a \in \Sigma_k} \#(a,w) \log_k g(r)(a)$$
$$= |w| \left( s - \sum_{a \in \Sigma_k} \theta_w(a) \log_k \frac{1}{g(r)(a)} \right)$$
$$= |w| \left( s - \mathrm{E}_{\theta_w} \log_k \frac{1}{g(r)(a)} \right)$$
$$= |w| \left( s - \mathrm{E}_{\theta_w} \log_k \frac{1}{\theta_w(a)} - \mathrm{E}_{\theta_w} \log_k \frac{\theta_w(a)}{g(r)(a)} \right)$$
$$= |w| \left( s - \mathcal{H}_k(\theta_w) - \mathcal{D}_k(\theta_w \parallel g(r)) \right).$$

We have thus shown that
$$d_{G_r}^{(s)} = k^{(s - \mathcal{H}_k(\theta_w) - \mathcal{D}_k(\theta_w \parallel g(r)))|w|} \tag{4.22}$$

holds for all $r \in \mathbb{Q} \cap [0,1]$, $s \in [0, \infty)$, and $w \in \Sigma_k^+$.

We now note a useful property of the function $g$. If we fix $r \in (0,1]$, then
$$\frac{d}{dx}[\mathcal{H}_k(g(x)) + \mathcal{D}_k(g(x) \parallel g(r))] = \frac{k-1}{k} \log_k \frac{k+r-kr}{r} > 0,$$

so
$$q \leq r \Rightarrow \mathcal{H}_k(g(q)) + \mathcal{D}_k(g(q) \parallel g(r)) \leq \mathcal{H}_k(g(r)). \tag{4.23}$$

For each $n \in \mathbb{Z}^+$, let $\theta_n^A = \theta_{w_n}$, where $w_n = \mathrm{CE}_k(A)[0..n-1]$ is the string consisting of the first $n$ symbols in $\mathrm{CE}_k(A)$. Then $\theta_1^A, \theta_2^A, \ldots$ is an infinite sequence of probability vectors in the simplex $\Delta(\Sigma_k)$. For every $n$ such that $T^{-1}(n)$ is even, $A_{=n} = C_{=n}^*$ consists entirely of integers $a$ for which $\theta_{\sigma_k(a)} = \vec{\pi}^{(n)}$, and for every $n$ such that $T^{-1}(n)$ is odd, $A_{=n} = D_{=n}^*$ consists entirely of integers $a$ for which $\theta_{\sigma_k(a)} = \vec{\tau}^{(n)}$. Since $\vec{\pi}^{(n)}$ converges to $g(r_\gamma)$, $\vec{\tau}^{(n)}$ converges to $g(r_\delta)$, and $G$ grows very rapidly, it follows easily that the set of limit points of the sequence $\theta_1^A, \theta_2^A, \ldots$ is precisely the closed line segment $g([r_\gamma, r_\delta])$ (which is a point if $\gamma = \delta$).

To see that $\dim_{\mathrm{FS}}(\mathrm{CE}_k(A)) \leq \gamma$, assume that $\gamma < s \leq 1$. It suffices to show that $\dim_{\mathrm{FS}}(\mathrm{CE}_k(A)) \leq s$. For this, fix $r \in \mathbb{Q} \cap (r_\gamma, (\mathcal{H}_k \circ g)^{-1}(s))$. Since $g(r_\gamma)$ is a limit point of $\theta_1^A, \theta_2^A, \ldots$, there is a sequence $n_1 < n_2 < \cdots$ of positive integers such that $\lim_{i\to\infty} \theta^A_{n_i} = g(r_\gamma)$. By (4.22), (4.23), and the continuity of $\mathcal{H}_k(\vec{x}) + \mathcal{D}_k(\vec{x} \parallel g(r))$ as a function of $\vec{x}$, we then have

$$\begin{aligned} d^{(s)}_{G_r}(w_{n_i}) &= k^{(s-\mathcal{H}_k(\theta^A_{n_i})-\mathcal{D}_k(\theta^A_{n_i}\|g(r)))n_i} \\ &= k^{(s-\mathcal{H}_k(g(r_\gamma))-\mathcal{D}_k(g(r_\gamma)\|g(r))-o(1))n_i} \\ &\geq k^{(s-\mathcal{H}_k(g(r))-o(1))n_i} \end{aligned}$$

as $i \to \infty$. Since $\mathcal{H}_k(g(r)) < s$, it follows that $G_r$ $s$-succeeds on $\mathrm{CE}_k(A)$, whence $\dim_{\mathrm{FS}}(\mathrm{CE}_k(A)) \leq s$.

To see that $\mathrm{Dim}_{\mathrm{FS}}(\mathrm{CE}_k(A)) \leq \delta$, assume that $\delta < s \leq 1$. It suffices to show that $\mathrm{Dim}_{\mathrm{FS}}(\mathrm{CE}_k(A)) \leq s$. For this, fix $r \in \mathbb{Q} \cap (r_\delta, (\mathcal{H}_k \circ g)^{-1}(s))$. For each $n \in \mathbb{Z}^+$, let $g(q_n)$ be the point on the line segment $g([r_\gamma, r_\delta])$ that is closest to $\theta^A_n$. Since $g([r_\gamma, r_\delta])$ contains every limit point of $\theta^A_1, \theta^A_2, \ldots$, $\Delta(\Sigma_k)$ is compact, and $\mathcal{H}_k(\vec{x}) + \mathcal{D}_k(\vec{x} \parallel g(r))$ is a continuous function of $\vec{x}$, we have

$$\mathcal{H}_k(\theta^A_n) + \mathcal{D}_k(\theta^A_n \parallel g(r)) = \mathcal{H}_k(g(q_n)) + \mathcal{D}_k(g(q_n) \parallel g(r)) + o(1) \tag{4.24}$$

as $n \to \infty$. By (4.22), (4.23), and (4.24),

$$\begin{aligned} d^{(s)}_{G_r}(w_n) &= k^{(s-\mathcal{H}_k(\theta^A_n)-\mathcal{D}_k(\theta^A_n\|g(r)))n} \\ &= k^{(s-\mathcal{H}_k(g(q_n))-\mathcal{D}_k(g(q_n)\|g(r))-o(1))n} \\ &\geq k^{(s-\mathcal{H}_k(g(r))-o(1))n} \end{aligned}$$

as $n \to \infty$. Since $\mathcal{H}_k(g(r)) < s$, it follows that $G_r$ strongly $s$-succeeds on $\mathrm{CE}_k(A)$, whence $\mathrm{Dim}_{\mathrm{FS}}(\mathrm{CE}_k(A)) \leq s$. □

Finally, we note that the Copeland-Erdös theorem is a special case of our main theorem.

**Corollary 4.3.** (Copeland and Erdös [8]). *Let $k \geq 2$ and $A \subseteq \mathbb{Z}^+$. If, for all $\alpha < 1$, for all sufficiently large $n \in \mathbb{Z}^+$, $|A \cap \{1, \ldots, n\}| > n^\alpha$, then the sequence $\mathrm{CE}_k(A)$ is normal over the alphabet $\Sigma_k$. In particular, the sequence $\mathrm{CE}_k(\mathrm{PRIMES})$ is normal over the alphabet $\Sigma_k$.*

*Proof.* The hypothesis implies that $\dim_\zeta(A) \geq \alpha$ for all $\alpha < 1$, i.e., that $\dim_\zeta(A) = 1$. By Theorem 4.2, this implies that $\dim_{\mathrm{FS}}(\mathrm{CE}_k(A)) = 1$, which is equivalent [16, 5] to the normality of $\mathrm{CE}_k(A)$. □